\documentstyle[12pt]{article}
\def\beq{\begin{equation}}
\def\eeq{\end{equation}}
\def\e{\epsilon}
\def\m{\mu}
\def\et{{\tilde \epsilon}}
\def\mt{{\tilde \mu}}

\begin{document}

\begin{titlepage}
\begin{center}
{\Large \bf Theoretical Physics Institute \\
University of Minnesota \\}  \end{center}
\vspace{0.3in}
\begin{flushright}
TPI-MINN-92/31-T \\
September 1993
\end{flushright}
\vspace{0.4in}
\begin{center}
{\Large \bf Catalyzed decay of false vacuum in four dimensions \\}
\vspace{0.2in}
{\bf M.B. Voloshin  \\ }
Theoretical Physics Institute, University of Minnesota \\
Minneapolis, MN 55455 \\
and \\
Institute of Theoretical and Experimental Physics  \\
                         Moscow, 117259 \\
\vspace{0.2in}
{\bf   Abstract  \\ }
\end{center}

The probability of destruction of a metastable vacuum state by the field
of a highly virtual particle with energy $E$ is calculated
for a (3+1) dimensional theory in the leading
WKB approximation in the thin-wall limit. It is found that the induced
nucleation rate of bubbles, capable of expansion, is exponentially small at
any energy. The negative exponential power in the rate reaches its maximum
at the energy, corresponding to the top of the barrier in the bubble energy,
where it is a finite fraction of the same power in the probability of the
spontaneous decay of the false vacuum, i.e. at $E=0$.

\end{titlepage}

A number of problems in statistical physics$^{\cite{lk,pp}}$ and in
cosmology$^{\cite{vko,linde}}$ involve a consideration of a metastable
(false) vacuum state of quantum fields, which corresponds to a local, rather
than global minimum of the Hamiltonian. Such state can
spontaneously decay into either the true vacuum or a lower-energy false
vacuum due to quantum fluctuations at zero
temperature$^{\cite{lk,pp,vko,coleman,cc}}$ or due to thermal
ones$^{\cite{langer1,langer2}}$ if the temperature is sufficiently high.

The decay proceeds through nucleation and subsequent expansion of bubbles
filled with the lower-energy phase. The expansion is possible only for
bubbles of sufficiently large size, for which the gain in the volume energy
compensates the energy associated with the surface of the bubble. Thus the
problem of calculation of the decay rate is reduced to a calculation of the
probability of nucleation of the critical bubbles, which
in the quantum case is a tunneling process$^{\cite{lk,pp,vko}}$.
The rate of the spontaneous nucleation of critical bubbles due to tunneling
is exponentially small in the inverse of the difference $\e$ of the energy
density between the metastable vacuum and the lower one. Thus it is
especially interesting to look for mechanisms, which would enhance the decay
rate.

If there are particles present in the false vacuum, they can facilitate
nucleation of the bubbles thus catalyzing the decay process. The presence of
a massive particle is known$^{\cite{adl,vs1}}$ to enhance the tunneling rate,
since the tunneling proceeds at energy equal to the particle mass rather
than zero, whereas the problem of the catalysis of the false vacuum decay by
collisions of particles thus far has been addressed either only for theories
in two dimensions$^{\cite{vs2,kiselev,rst}}$, or purely
phenomenologically$^{\cite{vko,els}}$.

In this paper is calculated for a (3+1) dimensional theory the exponential
power $-F(E)$ in the probability of the nucleation of critical and
subcritical bubbles in the presence of a highly virtual field $\phi$: $\left
| \langle B(E) | \phi | 0 \rangle \right |^2 \sim \exp(-F(E))$, with $|B(E)
\rangle$ being a state of a bubble with energy $E$. The calculations are
done within the so-called thin wall approximation, which assumes that the
size of the bubbles is much larger that the thickness of its wall and which
is applicable at small $\e$.  The result of this calculation is that the
induced nucleation rate of critical bubbles is exponentially small in
$\e^{-1}$ at any energy $E$.
The probability reaches its maximum at the
value of energy $E_c$ corresponding to the top of the barrier, which
separates the critical and subcritical regions. However at that point the
factor $F$ in the exponent differs only numerically from that at $E=0$.
The value of the ratio is found to be $F(E_c)/F(0) \approx 0.160$\,.
This behavior is different from the one derived$^{\cite{kiselev,rst}}$ for a
two-dimensional theory, where the exponential suppression in $\e^{-1}$
disappears at and above the top of the barrier, leaving only a possible
exponential suppression in the inverse of a coupling constant $g$ in the
theory: $\exp (- {\rm const}/g)$. As will be shown, the leading contribution
to the critical bubble nucleation rate at energy below the top of the
barrier is a product of two factors:  one being the probability of
excitation of a subcritical bubble with energy $E$ and the other given by
the tunneling rate at the same energy. At the top of the barrier the
suppression due to the tunneling disappears, however the excitation factor
is already exponentially small.  The difference with the two-dimensional
case arises from the fact that in the two-dimensional problem there is no
subcritical region for the bubbles in the thin-wall approximation (the
barrier starts at zero size of the bubble), hence the excitation factor
there is not related to the parameter $\e^{-1}$, but rather, possibly, to
$g^{-1}$.

The problem under discussion in this paper is closely related to the one of
multi particle production in high energy collisions in theories with weak
interaction (for a recent review see e.g. {\cite{mattis}}). Like some of the
recent papers on that subject$^{\cite{v,pg,ct,dp,gv}}$ the present
calculation uses the Landau-WKB technique$^{\cite{lwkb,ip}}$ for evaluating
matrix elements between strongly different states of a quantum system.

The simplest model, in which there is a false vacuum state, is the theory of
one real scalar field $\phi$ with the Lagrangian

\beq
{\cal L} = {1 \over 2} (\partial_\m \phi)^2 - {\lambda \over 4} (\phi^2 -
v^2)^2 - a\, \phi
\label{lagr}
\eeq
with $\lambda , \, v$ and $a$ being constants. In the limit of vanishing
asymmetry parameter $a$ the field has two degenerate vacuum states,
corresponding to $\langle 0 | \phi | 0 \rangle = \pm v$, For small positive
$a$ the state $\phi_+$ at $+v$ becomes a local minimum (false vacuum) and
the one near $-v$ $(\phi_-)$ becomes the true vacuum. The difference
$\e$ in the energy density between these states is given by

\beq
\e \equiv \e(\phi_+) - \e(\phi_-) =2\,a\,v + O(a^2)~~.
\eeq

The bubbles in the false vacuum are droplets of the phase $\phi_-$ embedded
in the phase $\phi_+$. The transition region between the phases (bubble
wall) is of the thickness $\sim 1/(\sqrt{\lambda} v)$, and throughout this
paper only the bubbles, whose characteristic size is much larger than this
scale, will be considered (thin wall approximation). The energy $E$ of a
bubble, as measured in the false vacuum, consists of a negative
part proportional to its volume: $-\e \, V$ and a positive part, associated
with the surface energy density $\m$. For small asymmetry parameter $a$ the
surface density can be taken as that of the domain wall in the symmetrical
limit

\beq
\m={2 \over 3} \sqrt{2 \lambda} v^3~~.
\eeq
In the tunneling process the lowest-action path is provided by spherical
bubbles, which have the maximal volume to surface ratio. Thus in the leading
WKB approximation it is sufficient to consider only spherically symmetrical
bubbles, whose dynamics in the thin-wall approximation is described in terms
of only one collective variable: the radius $r$. The classical
equations of motion are
determined by the following relation$^{\cite{vko}}$ for the Hamiltonian $H$

\beq
(H+\et r^3)^2-p^2= (\mt r^2)^2~~,
\label{ham}
\eeq
where $p$ is the canonical momentum conjugate of $r$ and
the notation $\et = {{4 \pi} \over 3} \e$ and $\mt = 4 \pi \, \m$ is
introduced in order to minimize the appearance of factors of $\pi$ in
subsequent formulas.

According to eq.(\ref{ham}) the potential energy of a bubble is given by the
sum of the (negative) volume term and the (positive) surface term:

\beq
U(r) \equiv H(r,\, p=0) = \mt r^2 - \et r^3~.
\label{upot}
\eeq
Thus, as shown in Figure 1, at an energy $E$ such that $0 < E < E_c = {4
\over {27}} (\mt^3 / \et^2)$ there are two classically allowed
regions for a bubble with energy $E$: the subcritical region to the left of
the barrier and the critical region to the right of the barrier. The bubbles
in the subcritical region oscillate and relatively
slowly$^{\cite{kobzarev}}$ dissipate their energy by emission of particles.
The bubbles in the critical region infinitely expand thus destroying the
false vacuum. At energy above $E_c$ there is no distinction between the
subcritical and critical bubbles, and nucleation of a bubble with such
energy would automatically imply destruction of the false vacuum.

A semi-classical quantization of
the effective theory with the Hamiltonian determined by eq.(\ref{ham})
enables one to calculate the rate of the spontaneous decay of the false
vacuum$^{\cite{vko}}$, and the same approach is used in what follows to
calculate the matrix elements $\langle B(E) | \phi | 0 \rangle$ by means of
the Landau-WKB technique. According to Landau$^{\cite{lwkb,ip}}$ for a
system with the coordinates $q$ the matrix element of an operator $f(q)$
between two strongly different states $| X(E_1) \rangle$ and $| Y(E_2)
\rangle$ with energies $E_1$ and $E_2$: $\langle Y(E_2) | f | X(E_1)
\rangle$  in the leading WKB  approximation is given by

\beq
|\langle Y(E_2) | f | X(E_1) \rangle| \sim
\exp~ \left [{\rm Re} \left ( i \int_{q_X}^{q^\ast} p(q;E_1)\,dq +
i \int^{q_Y}_{q^\ast} p(q;E_2)\,dq \right ) \right ]~~,
\label{landau}
\eeq
where $q^\ast$ is the (generally complex) `transition point', i.e. the
point of stationary phase of the expression

\beq
\exp \left ( i \int_{q_X}^{q^\ast} p(q;E_1)\,dq +
i \int^{q_Y}_{q^\ast} p(q;E_2)\,dq \right )~~,
\label{intd}
\eeq
$p(q,E_1)$ $(p(q,E_2))$ are the momenta on the classical (generally complex)
trajectory with energy $E_1$ $(E_2)$, which runs between the points
$q_X$ and $q^\ast$ ($q^\ast$ and $q_Y$), and, finally, $q_X$ and $q_Y$ are
points, chosen somewhere in the classically allowed regions for the states
$X$ and $Y$ correspondingly. The particular choice of each of the latter
points in a simply connected domain of the classically allowed region does
not affect the real part of the integrals in eq.(\ref{landau}). The
interpretation of the Landau formula (\ref{landau}) is straightforward
within the approach consistently pursued in the Landau-Lifshits textbook in
connection with the WKB calculation of various transition amplitudes: the
amplitude is given by the exponent of the truncated classical action on the
trajectory, which runs from the initial state to the final through a
(complex) `transition point'.

Few remarks are in order in connection with the
application of eq.(\ref{landau}) in the problem discussed here. First is
that eq.(\ref{landau}) is written for the case, relevant to present
calculation, when
the classical value of the operator $f$ is not
exponential at the `transition point' $q^\ast$, so that the
exponential factor, given by eq.(\ref{landau}) is not sensitive to the
specific form of the operator. Second is that eq.(\ref{landau}) does not
require the WKB approximation to be applicable for the wave functions of
either of the states $X$ and $Y$ in the classically allowed region, i.e.
where these wave functions are large.  Thus it can be applied even if the
lowest of the two energies, say $E_1$, is small, including the case $E_1=0$.
The only condition for applicability of eq.(\ref{landau}) is that the states
$X$ and $Y$ are `strongly different' in the sense that the matrix element,
given by this equation, contains large exponential power, i.e. that it is
strongly exponentially suppressed.  Third is that the branch of the function
$p(q,E)$ in the complex plane is to be chosen so that the exponential power
in eq.(\ref{landau}) is negative.  Finally, if there are several `transition
points' $q^\ast$, only the contribution of the one which gives the maximal
transition probability is to be retained.

In the matrix element $\langle B(E) | \phi | 0 \rangle$ the field operator
with zero spatial momentum (c.m. system) translates in the effective theory
of the thin-wall bubbles into the operator

\beq
\int (\phi(x) - \phi_+) \, d^3 x = {8 \over 3} \, \pi \,v\, r^3~~.
\label{transl}
\eeq
Thus the whole problem can be reformulated in terms of the effective
theory as a calculation of the matrix element $\langle B(E) | r^3 | 0
\rangle$ for a system with the Hamiltonian determined by eq.(\ref{ham}).
Using the Landau formula (\ref{landau}) one can write the exponential
estimate for this matrix element as

\begin{eqnarray}
|\langle B(E) | \phi | 0 \rangle| &\sim& \exp \left [ -{\rm Re} \left (
\int_0^{r^\ast} \sqrt{(\mt \, r^2)^2 - (\et \, r^3)^2} \, dr +
\int_{r^\ast}^{r(E)} \sqrt{(\mt \, r^2)^2 - (\et \, r^3+E)^2} \, dr \right )
\right ] \nonumber \\
&=& \exp \left [ -{1 \over \xi}{\rm Re} \left (
\int_0^{x^\ast} \sqrt{x^4-x^6} \, dx +
\int_{x^\ast}^{x(E)} \sqrt{x^4 - (x^3+w)^2} \, dx \right )
\right ]~,
\label{wf}
\end{eqnarray}
where instead of $r$ and $E$ the dimensionless variables $x$ and $w$ are
introduced as $ r = x\, \mt / \et$ and $E= w\, \mt^3 / \et^2$ and $\xi =
\et^3 / \mt^4$ is the small dimensionless constant in the effective theory
of bubbles. In Figure 2 are shown the classical turning points for bubbles at
zero energy and also for an energy $E < E_c$.  At $E=0$ the classically
allowed domain consists of the region $x > 1$ and of the point $x=0$.
At a non-zero energy $E < E_c$ the classically allowed domain consists of
two finite regions: to the left of barrier, $x < x_1(E)$, corresponding to
subcritical bubbles and to the right of the barrier, $x> x_2(E)$, which
corresponds to infinitely expanding critical bubbles. Accordingly the final
point $x(E)$ of the transition trajectory in eq.(\ref{wf}) can be chosen
either in the subcritical domain (path I~+~II in Fig. 2) or in the critical
one (path I~+~III in Fig. 2).  The former choice produces the amplitude of
the excitation of a subcritical bubble:  $A_- = \langle B_{\rm sub-c}(E)
|\phi |0 \rangle$, while the latter choice gives the amplitude of production
of an infinitely expanding critical bubble $A_+ = \langle B_{\rm c}(E) |\phi
|0 \rangle$. In either case the transition path starts at the point $x=0$
and with $E=0$, which corresponds to absence of a bubble in the initial
state.  Strictly speaking, the thin-wall approximation is not applicable at
$r=0$. However, the inaccuracy of the approximation at the values of
the radius of the order of the thickness of the wall does not affect the
factors $ \sim \exp( - {\rm const}/ \xi)$ which are being considered in this
calculation. In other words, the expression in eq.(\ref{wf}) receives
dominant contribution from the region of large $r$, and therefore is
calculable within the thin-wall approximation. From the paths shown in Fig.2
it is clear that the amplitudes $A_+$ and $A_-$ are related as

\beq
|A_+|=|A_-|\, \exp(-b(E)/\xi)~~,
\label{fact}
\eeq
where

\beq
b(E)/\xi=\int_{r_1(E)}^{r_2(E)} |p(r;\, E)|\, dr  =
{1 \over \xi} \int_{x_1(E)}^{x_2(E)} \sqrt{x^4-(x^3+w)^2} \,dx
\label{bar}
\eeq
is the exponential power in the barrier penetration rate at energy
$E$. The relation (\ref{fact}) can thus be interpreted as stating that the
production of the critical bubble at $E < E_c$ proceeds through excitation
of a subcritical one with subsequent tunneling through the barrier.

According to the expression (\ref{intd}) the `transition point' $x^\ast$ is
determined by solution of the equation

\beq
\sqrt{x^4-x^6}-\sqrt{x^4-(x^3+w)^2} = 0~~.
\label{xast}
\eeq
The solutions to this equation are given by the three
values of the cubic root $(-w/2)^{1/3}$. A simple inspection shows that
as the appropriate `transition point' one can choose either of the
complex values of the root in the right half plane (choosing one instead of
another gives the same result after proper redefinition of the branches of
the expressions in eq.(\ref{wf})). The integrals in eq.(\ref{wf}) were
evaluated numerically to determine the functions $c(E)$ and $b(E)$,
appearing in the amplitudes $A_-$ and $A_+$:

\beq
|A_-| \sim \exp \left (-{{c(E)} \over \xi}\right )~, ~~~~~~~~~~~~
|A_+| \sim \exp \left ( -{{c(E)+b(E)} \over \xi} \right ).
\label{bp}
\eeq
The results of the numerical calculation are shown in Fig. 3. At the
critical energy $E_c$, corresponding to the top of the barrier, the barrier
penetration term $b(E)$ vanishes. However the excitation term $c(E)$ at
this energy has a finite value $c(E_c) \approx  0.0314 \approx 0.160 \,
b(0)$, where $b(0)=\pi/16$ is the value of the barrier penetration term for
the spontaneous false vacuum decay. (In fact  $c(E_c) $ can be found
exactly in terms of elliptic integrals, but the final expression for the
result is unusually cumbersome.)

The function $c(E)$ can be found analytically in the limit of large $w$
as well as of small $w$.  For large $w$ one can neglect $x^4$ in comparison
with $x^6$ and with $(x^3+w)^2$ in eq.(\ref{wf}) and thus find

\beq
c(E) = {{3 \sqrt{3}} \over 4} \left | {w \over 2} \right |^{4 \over 3}~~~
{}~~~~(w \gg 1)~.
\label{wlarge}
\eeq
For $w \ll 1$ the expression in eq.(\ref{wf}) is determined by the region of
$x$ near the classical turning point $x_1(E)$. In this region one can
neglect in eq.(\ref{wf}) $x^6$ in comparison with $x^4$ and also neglect
$x^3$ in comparison with $w$. Then $c(E)$ can be found as

\beq
c(E)= \int_0^L x^2 \,dx- \int_{\sqrt{w}}^L \sqrt{x^4-w^2} \,dx =
{\sqrt{\pi} \over 6} {{\Gamma (1/4)} \over {\Gamma (3/4)}} w^{3 \over 2}
\approx  0.874 w^{3 \over 2}~~~~~~~~(w \ll 1)~,
\label{wsmall}
\eeq
where both integrals run
along the real axis and $L$ is a cutoff parameter, $L \gg \sqrt{w}$.
The difference of the integrals is determined by the region $x
\sim \sqrt{w}$, which substantiates the approximation, leading from
eq.(\ref{wf}) to eq.(\ref{wsmall}). The full exponential power in the
excitation amplitude $A_-$ for small $w$ is thus given by $c(E)/\xi = {\rm
const} \, E \sqrt{E/\m}$ which coincides with the result for the
amplitude of excitation of a bubble with energy $E$ in the case of
degenerate vacua obtained in \cite{gv}.  (Clearly, in that case only
subcritical bubbles exist).
One should however keep in mind that the region of small $w$ is limited from
below by the condition of applicability of the thin-wall approximation,
which implies that the characteristic size of the bubbles in the relevant
region $r \sim \sqrt{w}\,\mt/\et$ is larger than the thickness of the
wall. In terms of $E$ this translates into the condition$^{\cite{gv}}$ $E
\gg \m^{1/3}$.

It can be also noticed that at small energy the barrier
penetration term

\beq
b(E)={\pi \over 16} - w +o(w)
\eeq
decreases faster than the $w^{3/2}$ growth of the $c(E)$. Therefore the
probability of the induced decay of the false vacuum grows with energy in
this region.  As is seen from Fig. 3, this behavior continues up to the top
of the barrier, where $b(E)$ vanishes.

The behavior of the induced tunneling amplitude calculated in this paper is
similar to the one observed$^{\cite{ct,dp}}$ in the quantum-mechanical
example with the double well potential $(x^2-1)^2$, where at the top of the
barrier the exponential power in the excitation probability is a finite
fraction, namely one half, of that in the tunneling probability at $E=0$.
That the ratio of the exponential powers in that case is exactly one half is
a consequence of the reflection symmetry of the potential and of the
standard relation of the Hamiltonian to the kinetic and the potential energy.
Both these features do not hold for the problem discussed here, hence the
particular value of the ratio of the exponential powers is different, and is
approximately equal to 0.160\,.

As a final remark one can note, that the Landau formula (\ref{landau}) is
not sensitive in the leading exponential approximation to the particular
form of the operator $f(q)$, provided that the function $f(q)$ by itself is
not exponential in the parameters in the problem. Therefore, though for
definiteness the catalysis of the false vacuum decay by the particular
operator $\phi$ has been discussed, the same results should be applicable
for destruction of the false vacuum in any few-particle process at energy
$E$. Also one can notice, that the particular form of the Lagrangian in
eq.(\ref{lagr}) was used only to give the parameters $\e$ and $\m$ a
particular expression in terms of the underlying theory. The rest of the
calculation is based on the relation (\ref{ham}) for the Hamiltonian of the
effective theory, which is a general relation for the dynamics of spherical
bubbles in the thin-wall approximation. Therefore the results of the present
calculation are applicable whenever the latter approximation is valid.

This work is supported in
part by the DOE grant DE-AC02-83ER40105.

{\Large \bf Figure captions}\\[0.15in]
{\bf Fig. 1}. Potential energy of a bubble vs. its radius.\\[0.1in]
{\bf Fig. 2}. Classical turning points and the transition path in the Landau
formula for the bubbles. $x=1$ is the turning point on the right
of the barrier at zero energy. $x_1$ and $x_2$ are the turning
points on the left and on the right of the barrier at energy $E$. The
transition trajectory starts at $x=0$ and goes with energy $E=0$ to the
`transition point' $x^\ast$ (the link I), then it goes with energy $E$
either to the subcritical region (the link II) or to the critical one
(the link III).\\[0.1in]
{\bf Fig. 3}. The barrier penetration function $b(E)$ (dashed), the
excitation function  $c(E)$ (dotted), and their sum (solid) vs.
$w=E\,\et^2/\mt^3$. At the point $w_c=4/27$ and beyond the barrier
disappears, hence $b(E)=0$ and the sum coincides with $c(E)$.

\newpage
\thispagestyle{empty}
\unitlength=1mm
\thicklines
\begin{picture}(130.00,120.00)
\put(10.00,50.00){\line(1,0){120.00}}
\put(20.00,35.00){\line(0,1){85.00}}
\put(20.00,50.00){\vector(3,4){16.50}}
\put(36.50,72.00){\line(3,4){13.50}}
\put(50.00,90.00){\vector(0,-1){20.00}}
\put(50.00,70.00){\line(0,-1){20.00}}
\put(50.00,90.00){\vector(4,-3){26.00}}
\put(76.00,70.50){\line(4,-3){27.33}}
\put(110.00,50.00){\circle*{1.50}}
\put(94.00,50.00){\circle*{1.50}}
\put(67.00,50.00){\circle*{1.50}}
\put(50.00,90.00){\circle*{1.50}}
\put(23.00,113.00){\makebox(0,0)[lc]{{\large Im\,$x$}}}
\put(122.00,47.00){\makebox(0,0)[ct]{{\large Re\,$x$}}}
\put(110.00,47.00){\makebox(0,0)[ct]{1}}
\put(67.00,47.00){\makebox(0,0)[ct]{{\large $x_1$}}}
\put(94.00,47.00){\makebox(0,0)[ct]{{\large $x_2$}}}
\put(50.00,93.00){\makebox(0,0)[cb]{{\large $x^\ast$}}}
\put(34.00,65.00){\makebox(0,0)[lt]{{\large I}}}
\put(52.00,65.00){\makebox(0,0)[lt]{{\large II}}}
\put(79.00,65.00){\makebox(0,0)[rt]{{\large III}}}
\put(65.00,00.00){\makebox(0,0)[cc]{{\bf Fig. 2}}}
\end{picture}

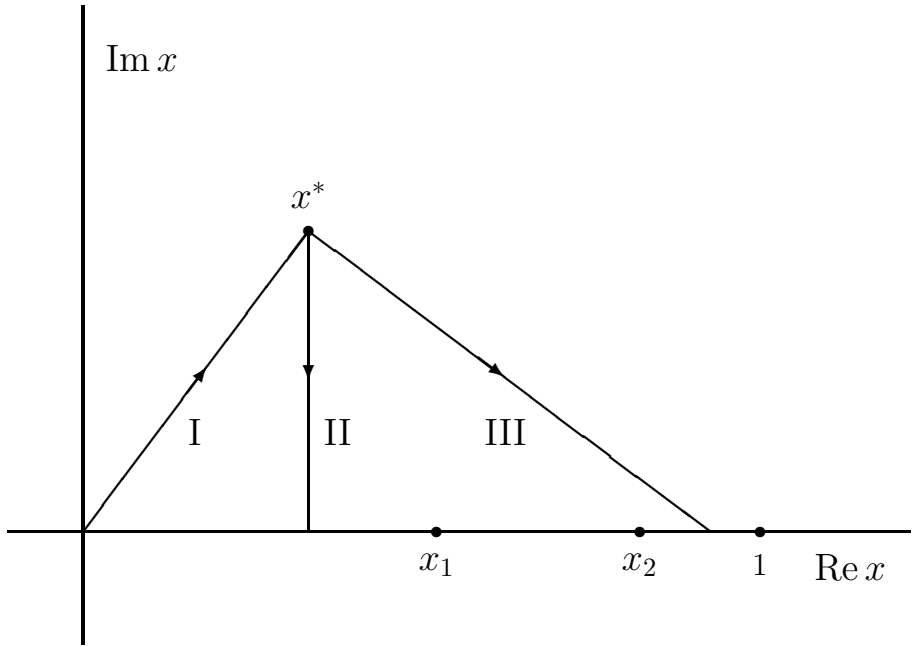
\end{document}